\documentclass[preprint,2p,onecolumn,numbers,sort&compress]{elsarticle}

\makeatletter
\protected\def\xvcenter{%
  \hbox\bgroup$\everyvbox{\everyvbox{}\aftergroup\m@th\aftergroup$\aftergroup\egroup}%
  \vcenter
}
\DeclareRobustCommand{\midscript}[1]{
  \mathchoice{\mid@script\scriptstyle{#1}}
    {\mid@script\scriptstyle{#1}}
    {\mid@script\scriptscriptstyle{#1}}
    {\mid@script\scriptscriptstyle{#1}}
}
\newcommand{\mid@script}[2]{
  \vcenter{\hbox{$\m@th#1#2$}}
}
\DeclareRobustCommand{\textmidscript}[1]{%
  \xvcenter{\hbox{\scriptsize#1}}%
}
\makeatother
\def\appendixname{}
\makeatletter
\renewcommand\appendix{\par
  \setcounter{section}{0}%
  \setcounter{subsection}{0}%
  \setcounter{equation}{0}
  \gdef\thefigure{\@Alph\c@section.\arabic{figure}}%
  \gdef\thetable{\@Alph\c@section.\arabic{table}}%
  \gdef\thesection{\appendixname\@Alph\c@section}%
  \@addtoreset{equation}{section}%
  \gdef\theequation{\@Alph\c@section.\arabic{equation}}%
  \addtocontents{toc}{\string\let\string\numberline\string\tmptocnumberline}{}{}
}
\makeatother
\usepackage{pdflscape}
\usepackage{afterpage}
\usepackage{fancyhdr}
\fancypagestyle{lscape}{%
\fancyhf{} % clear all header and footer fields
\fancyfoot[LE]{}
\fancyfoot[LO] {}

}
\usepackage{rotating}
\usepackage{pdflscape}
\usepackage{eso-pic}
\usepackage{zref-user}
\usepackage{scalerel}
\usepackage[utf8]{inputenc}
\makeatletter
\newcounter{cntsideways}
\AddToShipoutPictureBG{%
 \ifnum\zref@extractdefault{rotate\number\value{page}}{page}{0}=0
  \PLS@RemoveRotate
 \else
  \PLS@AddRotate{90}%
 \fi}
\newcommand\rotatesidewayslabel{\stepcounter{cntsideways}%
 \zlabel{tmp\thecntsideways}\zlabel{rotate\zref@extractdefault{tmp\thecntsideways}{page}{0}}}
\makeatother
\usepackage{amsmath}
\usepackage{amssymb}
\usepackage{booktabs}
\usepackage[inline]{enumitem}
\usepackage{geometry}
\usepackage{hyperref}
\usepackage{cleveref}
\crefname{lstlisting}{listing}{listings}
\Crefname{lstlisting}{Listing}{Listings}
\crefname{appsec}{appendix}{appendices}
\Crefname{appsec}{Appendix}{Appendices}
\usepackage[nowatermark]{fixmetodonotes}
\usepackage{isotope}
\usepackage{verbatim}
\usepackage{siunitx}
\DeclareSIUnit\clight{\text{\ensuremath{c}}}
\DeclareSIUnit\atom{\text{atom}}
\DeclareSIUnit\ton{\text{t}}
\DeclareSIUnit\parsec{\text{pc}}
\usepackage{xcolor}
\definecolor{lightgray}{rgb}{.9,.9,.9}
\definecolor{darkgray}{rgb}{.4,.4,.4}
\definecolor{purple}{rgb}{0.65, 0.12, 0.82}
\definecolor{darkred}{rgb}{0.64, 0.0, 0.0}
\usepackage{listings}
\lstdefinelanguage{MARLEYConfig}{
  keywords={seed, structure, reactions, source, direction, log,
    executable_settings,},
  keywordstyle=\color{black}\bfseries,
  ndkeywords={type, neutrino, Emin, Emax, temperature, eta, x, y, z,
    file, level, overwrite, events, output, format, mode},
  ndkeywordstyle=\color{black}\bfseries,
  identifierstyle=\color{black},
  sensitive=false,
  comment=[l]{//},
  morecomment=[s]{/*}{*/},
  commentstyle=\color{blue}\ttfamily,
  stringstyle=\color{darkred}\ttfamily,
  escapechar=|,
  morestring=[b]"
}
\DeclareFontFamily{OT1}{pzc}{}
\DeclareFontShape{OT1}{pzc}{m}{it}{<-> s * [1.10] pzcmi7t}{}
\DeclareMathAlphabet{\mathpzc}{OT1}{pzc}{m}{it}
\usepackage{graphicx}
\usepackage{subcaption}
\usepackage[mathlines]{lineno}
\newcounter{bla}

\journal{Computer Physics Communications}
\begin{document}
\newcommand{\marley}{\texttt{MARLEY}}
\newcommand{\rootcern}{ROOT}
\newcommand{\ascii}{ASCII}
\newcommand{\hepevt}{HEPEVT}
\newcommand{\cpp}{C\textmidscript{++}}%prettier C++
\newcommand{\version}{1.2.0}
\newcommand{\minGCCversion}{4.9.4}
\newcommand{\minClangVersion}{3.5.2}
\newcommand{\myRepo}{\url{https://github.com/davidc1/TRANSLATE}}
\newcommand{\myDOI}{\href{http://doi.org/10.5281/zenodo.3905443}
{10.5281/zenodo.3905443}}
\newcommand{\totKinE}{\varepsilon}
\newcommand{\fragmentKinELab}{ \varepsilon_\text{lab} }
\newcommand{\totKinEmax}{\varepsilon_\text{max}}
\newcommand{\totWidth}{\Gamma}
\newcommand{\SMatrixElement}{\left< S_{\ell j} \right>}
\newcommand{\xmin}{a}
\newcommand{\xmax}{b}
\newcommand{\fmax}{f_\text{max}}
\newcommand{\myN}{N}
\newcommand{\fragment}{ \ensuremath{a} }
\newcommand{\matchThresh}{ V_\text{thresh} }
\newcommand{\matchScale}{ \mathcal{S} }
\newcommand{\pdgCodeVar}{ \mathrm{P_{ID}} }
\newcommand{\levelIndex}{ \Lambda }
\newcommand{\ExCut}{ E_x^{\,\text{cut}} }
\newcommand{\decayProductIndex}{ \ensuremath{u} }
\newcommand{\exitChannelIndex}{ \ensuremath{e} }
\newcommand{\termIndex}{ \ensuremath{b} }
\newcommand{\FluxAvgTotXSec}{ \left<\sigma\right> }
\newcommand{\FluxAvgDiffXSec}{ \left<\frac{d\sigma}{dx}\right> }
\newcommand{\fragmentMomCM}{ \mathpzc{k} }
% Final nucleus rest frame variables
\newcommand{\FNRpLep}{\mathpzc{K}}
\newcommand{\FNReLep}{\mathpzc{E}}
\newcommand{\FNRpLepEff}{\FNRpLep_\text{\,\,eff}}
\newcommand{\FNReLepEff}{\FNReLep_\text{\,eff}}

\begin{frontmatter}
\title{\texttt{TRANSLATE} - A Monte Carlo Simulation of Electron Transport in Liquid Argon}

%\author[a]{To Be Determined}

\author[a,c]{Z.~Beever}
\author[c,d]{D.~Caratelli\corref{corr}}
\cortext[corr] {Corresponding author.\\\textit{E-mail address:} dcaratelli@ucsb.edu}
\author[c]{A.~Fava}
\author[b]{F.~Pietropaolo}
\author[b,e]{F.~Stocker}
\author[c]{J.~Zettlemoyer}

% Institutions in alphabetical order
\address[a]{Boston University, Boston, MA, 02215, USA}
\address[b]{CERN, The European Organization for Nuclear Research, 1211 Meyrin, Switzerland}
\address[c]{Fermi National Accelerator Laboratory, Batavia, IL, 60510, USA}
\address[d]{University of California Santa Barbara, Santa Barbara, CA, 93106, USA}
\address[e]{Yale University, New Haven, CT, 06520, USA}

\begin{abstract}
The microphysics of electron and photon propagation in liquid argon is a key component of detector design and calibrations needed to construct and perform measurements within a wide range of particle physics experiments. As experiments grow in scale and complexity, and as the precision of their intended measurements increases, the development of tools to investigate important microphysics effects impacting such detectors becomes necessary. In this paper we present a new time-domain Monte Carlo simulation of electron transport in liquid argon. The simulation models the TRANSport in Liquid Argon of near-Thermal Electrons (\texttt{TRANSLATE}) with the aim of providing a multi-purpose software package for the study and optimization of detector environments, with a particular focus on ongoing and next generation liquid argon neutrino experiments utilizing the time projection chamber technology. \texttt{TRANSLATE} builds on previous work of Wojcik and Tachiya, amongst others, introducing additional processes, including ionization, thus modeling the full range of drift electron scattering interactions. The simulation is validated by benchmarking its performance with swarm parameters from data collected in experimental setups operating in gas and liquid.
\end{abstract}
% FERMILAB-PUB-22-893-ND

\end{frontmatter}
%\linenumbers
%\rightlinenumbers*
% Computer program descriptions should contain the following
% PROGRAM SUMMARY.
\noindent
{\bf PROGRAM SUMMARY} \\
\begin{small}
\noindent
{\em Program Title:} \texttt{TRANSLATE}\\
%{\em Program Files doi:} \myDOI\  \\
{\em Developer's repository link:} \myRepo\ \\
{\em Licensing provisions:} GNU General Public License 3.0 \\
{\em Programming language:} \cpp, python\\
{\em External routines/libraries used:}
make, cmake (required), ipython notebook (optional)\\
{\em Nature of problem:} simulation of electron-argon scattering events at
energies of $10^{-3} - 10^3$ eV\\
{\em Solution method:} Monte Carlo time-domain transport of electrons simulating momentum transfer, excitations, and ionization interaction modes utilizing input single and double-differential cross sections. \\
{\em Restrictions:} effects of ion space-charge or recombination with positive argon ions are not included.
\end{small}



\section{Introduction}

The microphysics of electron propagation in noble elements determines performance metrics in a wide-range of detector technologies employed in particle physics. A solid understanding of this microphysics, and the availability of simulation tools which can describe its behavior, can help assess detector performance and guide technology R\&D. In this work we present a new simulation package, \texttt{TRANSLATE}, aimed at modeling the transport of electrons in gaseous and liquid argon accounting for electron-argon interactions from $10^{-3}$ to $10^3$ eV. \texttt{TRANSLATE} focuses on tracking the 3D trajectories and energy of drifting electrons in conditions of position-dependent electric field, and can be leveraged for a number of studies which aim to characterize detector response. Such a simulation tool can find valuable application in the modeling of electron transport in different environmental conditions. 
%In this work we focus in particular on the application of the simulation to the study of the conditions needed to produce secondary electron ionization in a liquid argon environment. This focus is motivated by the possibility of achieving electron amplification in LAr in order to improve signal strength and measure low-yield interactions such as nuclear recoils in LAr detectors. %Future versions of this simulation which aim to incorporate additional microphysics processes such as $e-{\textrm{Ar}}^{+}$ cross sections and photon production will be leveraged for the study of ion recombination and scintillation light production, both key to numerous applications in neutrino physics with large-scale LArTPCs.

Software packages for the modeling of the bulk behavior of drifting electrons in noble gasses have been developed in the past, such as the~\href{https://magboltz.web.cern.ch/magboltz/}{\texttt{Magboltz}} software, which uses  Monte-Carlo integration for the propagation of electrons in gas mixtures~\cite{bib:MagboltzWWW}. A recently released simulation package, \texttt{PyBoltz}~\cite{bib:pyboltz}, expands on the \texttt{Magboltz} software and modernizes the code. In this work we expand on the simulation developed by Wojcik and Tachiya~\cite{bib:WT} to develop a Monte-Carlo simulation of the microphysics of electron transport in argon in the time-domain, accounting for scattering, excitation, and ionization interactions of electrons. The Monte-Carlo approach to this simulation allows for the implementation of variations in the local electric field on a microscopic scale, enabling the study of specific scenarios which present variability in detector conditions. This simulation has probed variations at the nanometer level. We furthermore choose to focus the work on modeling of electron propagation and interactions in liquid argon, for which microphysics simulation tools are more scarce. This effort is in part motivated by the broad range of neutrino experiment Time Projection Chamber (TPC) detectors such as ArgoNeut~\cite{bib:argoneut}, LArIAT~\cite{bib:lariat}, MicroBooNE~\cite{bib:microboone}, ICARUS~\cite{bib:icarus}, protoDUNE~\cite{bib:protodune}, and DUNE~\cite{bib:dune} operating in liquid argon. 

Calculations of cross sections for electron interactions with atoms are a core ingredient of this work. For interactions in gaseous media, detailed datasets are available through the \href{https://us.lxcat.net/home/}{\texttt{LxCAT}}~\cite{bib:lxcat,bib:LXCAT1,bib:LXCAT2,bib:LXCAT3} online database. Such datasets for interactions in liquid are more scarce. For scattering interactions in liquid argon the cross section model adopted by Wojcik and Tachiya~\cite{bib:WT} is used by default. Recent calculations by G. Boyle et al. provide double-differential theoretical calculations of electron-argon scattering cross sections in both gas and liquid phase~\cite{bib:boyle}, which have been incorporated in the simulation as optional cross section inputs. 

A specific goal for the developed simulation is the investigation of the conditions under which electron amplification can be sustained in liquid argon, an R\&D topic with significant implications for detector technology development in nuclear and particle physics. This focus is motivated in particular by the possibility of measuring low-yield interactions such as nuclear recoils in single-phase LAr detectors. The attempt to achieve electron amplification in liquid argon has been pursued in the past with mixed results~\cite{bib:larmult01,bib:larmult02,bib:larmult03,bib:larmult04}. 
A more general goal of this work is to provide the growing liquid argon community with a simulation tool capable of comprehensive calculations of electron transport in liquid argon for the study of detector physics and detector performance. Particular avenues of development which are being considered are the possibility of expanding the current simulation to include positive Ar ions for a full micro-physics treatment of ion recombination, and the inclusion of impurities and contaminants in the model for the study of electron quenching as well as secondary processes which can lead to photon production. 

\textcolor{black}{The \texttt{TRANSLATE} simulation package is intended to be used either as a stand-alone tool aimed at tracking individual or swarms of ionization electrons in different detector environments, or coupled to general purpose Monte Carlo codes as part of a detector's particular simulation of ionization produced by the passage of ionizing charged particles. In this second scenario, users will need to provide a list of electron positions and energies, as well as the detector local electric field. Electric field values can be provided either as a fixed value or in the form of a 3D map $\vec{E}(x,y,z)$. As a note because electrons are propagated individually, the run time for a full MC simulation will scale linearly with the number of ionization electron a user aims to propagate.}

The remainder of this paper is structured as follows: Chapter~\ref{sec:xsec} describes the interactions that drifting electrons undergo in argon and presents the cross section models used in this simulation. Chapter~\ref{sec:simulation} describes the technical implementation of the \texttt{TRANSLATE} simulation. Chapter~\ref{sec:swarm} presents validations of the simulation by benchmarking swarm parameters calculated with the simulation to measurements from data. Chapter~\ref{sec:ionization} presents studies of electron amplification in argon obtained with the simulation first in gas, where results are validated with data, and subsequently in liquid. Finally, chapter~\ref{sec:conclusions} presents concluding remarks and an outlook for future developments.

\section{Electron-Argon Cross Sections}
\label{sec:xsec}

\begin{figure}%[H]
\centering
  \includegraphics[width=0.95\textwidth]{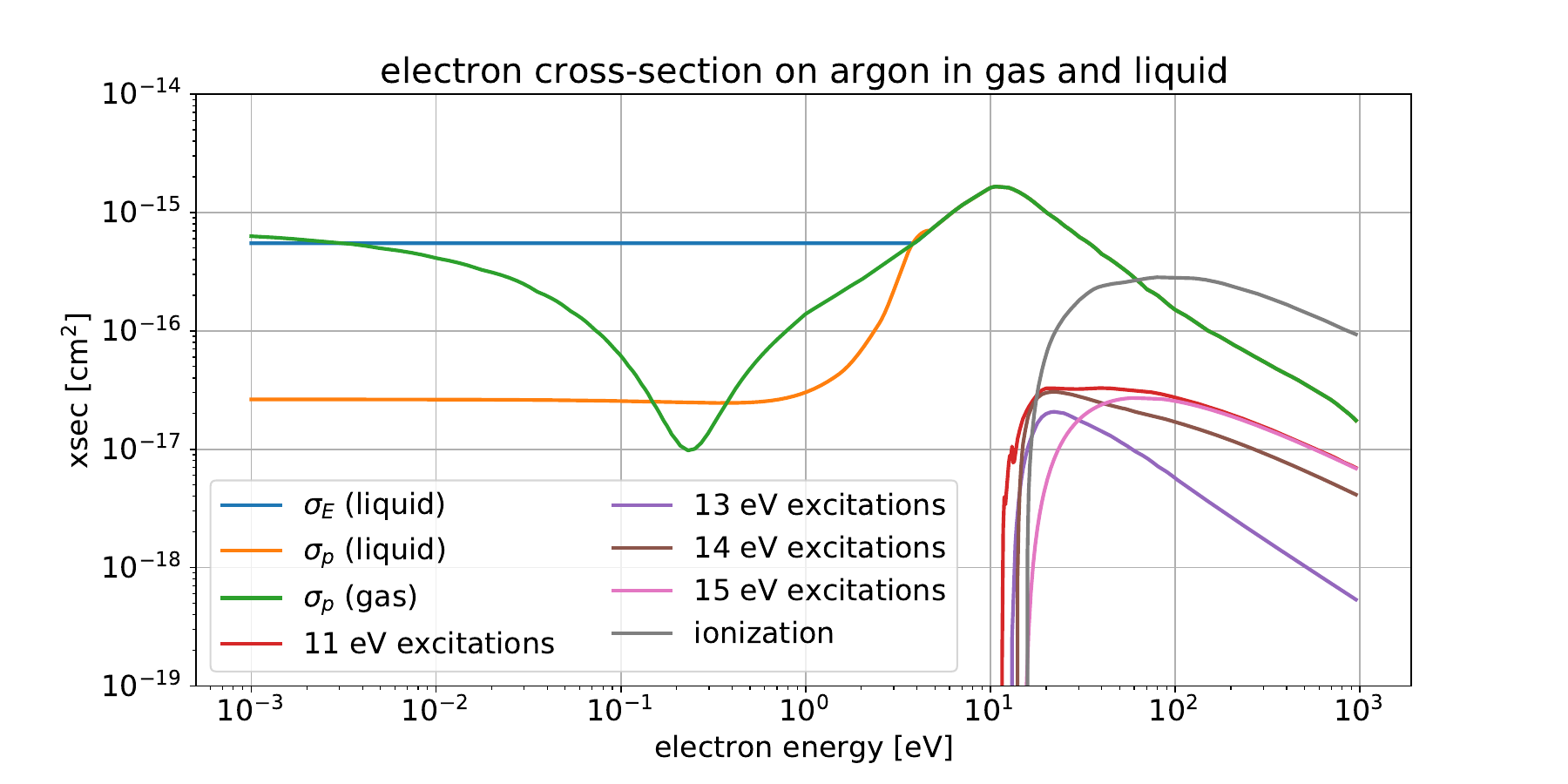}
  %\captionsetup{margin=0.1cm}
  \caption{cross section for electrons on argon atoms in the sub-eV to keV energy range. Data for gas is taken from the \texttt{LxCAT} database~\cite{bib:lxcat,bib:LXCAT1,bib:LXCAT2,bib:LXCAT3}. The cross section for liquid argon comes from Ref.~\cite{bib:WT}.}
  \label{fig:xsec1D}
\end{figure}

The interaction cross sections for drifting electrons on argon atoms is the primary physics input to the simulation. In order to adequately model electrons over a broad range of electric fields, we incorporate energy-dependent cross sections spanning six orders of magnitude, from $10^{-3}$ eV, below the electron's thermalization energy in liquid argon, to $10^{3}$ eV, where the cross section for ionization dominates the total interaction rate. The model accounts for elastic momentum-transfer and energy-transfer interactions, as well as inelastic excitations and ionization. 
%The former dominate the interaction cross section below several tens of eV. 
Figure~\ref{fig:xsec1D} shows the cross section for electrons scattering on argon as a function of the electron's kinetic energy for different interaction processes. For the elastic energy and momentum transfer processes, the different cross sections for gas and liquid argon are portrayed. The remainder of this chapter discusses each interaction mode in more detail, along with its implementation in the simulation.

\subsection{Elastic and Momentum-Transfer Scattering}
The dominant interaction mode at lower energies is associated with elastic scattering of electrons on argon. The treatment of this interaction process in the simulation differs in gas and liquid, and both are described here. 
%In this work we follow the simulation treatment developed and described in Ref.~\cite{bib:WT}. \textcolor{red}{expand on this section...} 
The cross section for elastic scatters in gas is taken from tabulated values from the \texttt{LxCat} database~\cite{bib:lxcat}. The strong suppression of the cross section in the region between 0.1 and 1 eV of electron energy is known as the Ramsauer-Townsend effect. \textcolor{black}{Modeling of the cross section in liquid is complicated by the dense medium of the liquid. The electron's mean free path in liquid argon is comparable to the de Broglie wavelength of electrons of sub-eV scale energy, leading to non-trivial quantum interference effects which cause such electrons to interact with correlated argon nuclei. Cohen and Lekner formulated a model which accounts for the complexity of scattering in a dense medium with correlated pairs of atoms~\cite{bib:CL}. This theory predicts different rates of energy and momentum transfer to electrons and provides distinct mean free paths for momentum and energy transfer, but does not present a microscopic description of the electron's transport. Wojcik and Tachyia first adapted the Cohen-Lekner theory (incorporating developments by  Atrazhev and Iakubov~\cite{bib:AI}) into a proper time-domain simulation by separating the elastic scattering interaction into two components associated with energy and momentum transfer, respectively. This work, documented in ~\cite{bib:WT} serves as the basis for this component of the \texttt{TRANSLATE} simulation.} Cross section for liquid are treated as described in Ref.~\cite{bib:WT}. %\sout{The Cohen-Lekner model serves as the basis theoretical calculation for the input cross section in liquid.}
\textcolor{black}{Following this work,} the elastic cross section \textcolor{black}{in \texttt{TRANSLATE}} is separated in two components: energy-transfer ($\sigma_E$, blue in Fig.~\ref{fig:xsec1D}) and momentum-transfer ($\sigma_p$, orange in Fig.~\ref{fig:xsec1D}). For elastic energy transfer collisions, energy transfer is computed classically assuming scattering off of an argon atom vith velocity sampled from its Boltzmann distribution. These collisions modify the electron's total velocity, but not its direction. As discussed in Ref.~\cite{bib:WT}, taking from the model developed by Atrazhev and Iakubov~\cite{bib:AI}, these collisions can be considered as scattering of electrons on ``interatomic vibrations (phonons)''. The momentum scattering cross section is modeled as elastic collisions which lead to an isotropic deflection of the outgoing electron. The values of $\sigma_E$ and $\sigma_p$ are taken from~\cite{bib:AI}, as described in Ref.~\cite{bib:WT}. At higher electron energies, where the wavelength of the electron becomes smaller than inter-atomic distances, the cross section for liquid becomes equal to that in gas. 

\subsection{Excitations}
Above $\mathcal{O}$(10) eV, enough energy is available to excite the argon atom. Over forty different excitation cross sections are documented in the \texttt{LxCAT} database from the Biagi  database~\cite{bib:biagiexcitation}. While all are accounted for in this simulation, for bookkeeping purposes they are grouped in~$\sim1$ eV bands near 11, 13, 14, and 15 eV. While an important contribution to the total interaction cross section, excitations have a significantly lower cross section compared to elastic interactions. In the simulation, whenever an excitation interaction occurs, an appropriate amount of energy is subtracted from the electron. \textcolor{black}{The direction of the original electron causing the excitation is however left unchanged.} Depending on the simulated excitation process, the energy subtracted from the electron is 11.68, 13.21, 14.10, or 15.23 eV. These values cover the span of excitation energies but do not fully encompass the complexity of individual levels that can be excited. To address the impact of this approximation we have studied the implementation of the modeling of excitations by assuming all have a single energy between 11 and 15 eV, and have found no noticeable variation in the observables of drift velocity, diffusion, or amplification studied in this work. Therefore, we conclude that the simulation is not particularly sensitive to the detailed treatment of the energy levels of the excitations.

\textcolor{black}{Not currently considered by the simulation is the production of Auger electrons and fluorescence photons by the relaxation of atomic excitations. While the simulation of these is not expected to considerably alter the results of the simulation, their inclusion is an area of improvement worth exploring in future work.}

\subsection{Ionization}

Above $\sim11$ eV, drifting electrons carry enough energy to knock out an electron orbiting the struck argon atom. When this occurs, a secondary free electron is liberated and begins to drift under the action of the external electric field, contributing to forming an ionization electron cloud. The cross section for ionization in this simulation comes from Ref.~\cite{bib:ionizationxsec1,bib:ionizationxsec2}, and is shown in black in Fig.~\ref{fig:xsec1D}. In the simulation, when an ionization interaction occurs, the ionization energy of 15.76 eV is subtracted from the energy of the drifting electron, and a secondary electron is produced with an initial energy sampled uniformly in the range 1 to 5 eV.

\subsection{Differential Cross Sections}

\begin{figure}%[H]
    \centering
    \begin{subfigure}[b]{0.45\textwidth}
    \centering
    \includegraphics[width=1.00\textwidth]{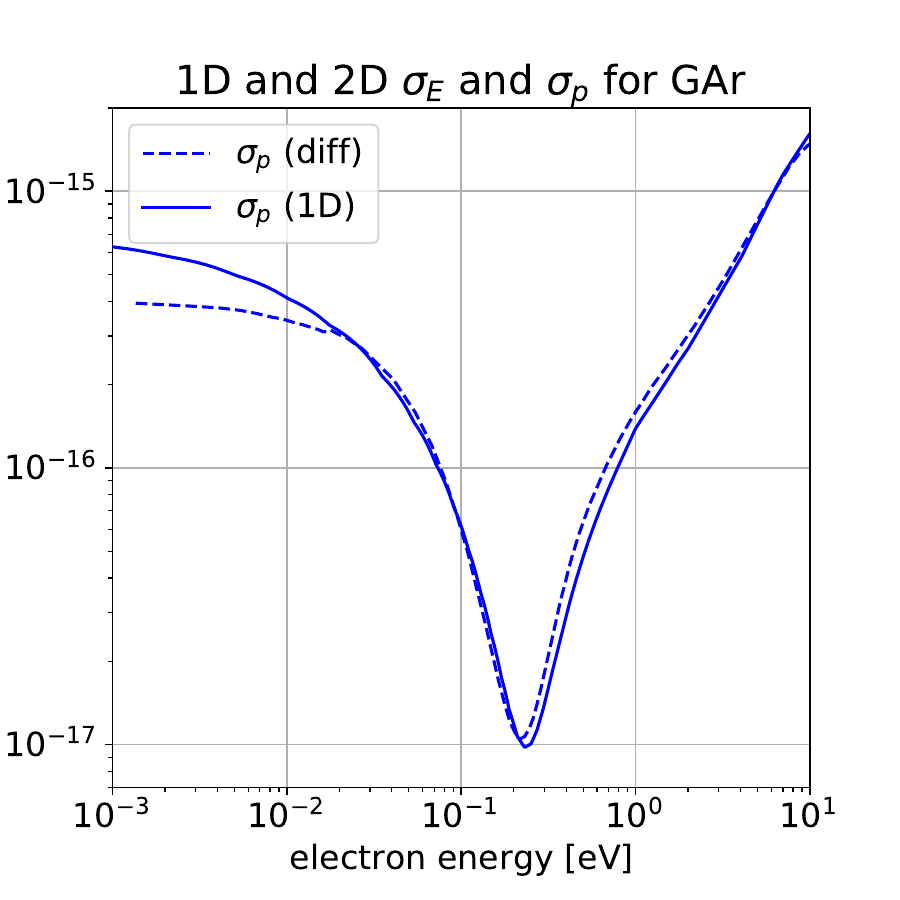}
    \caption{Cross section in gaseous argon}
    \end{subfigure}
    \begin{subfigure}[b]{0.45\textwidth}
    \centering
    \includegraphics[width=1.00\textwidth]{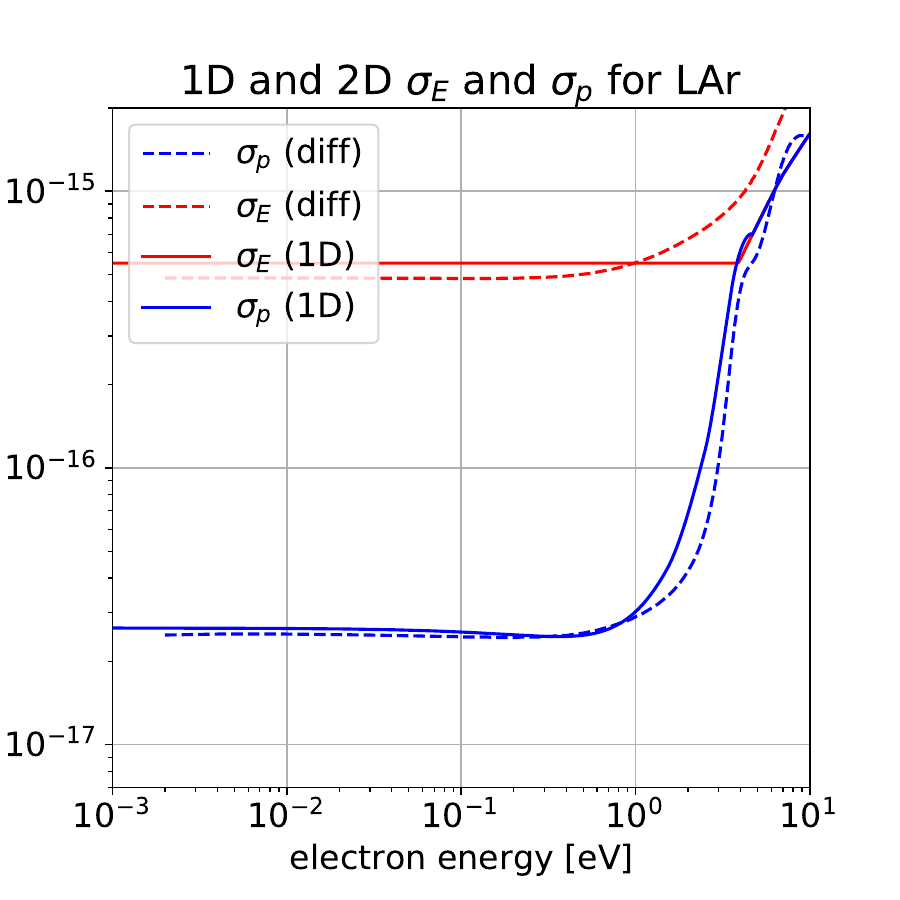}
    \caption{Cross section in liquid argon}
    \end{subfigure}
    \caption{Comparison of integrated double-differential cross sections (dashed lines) from the \texttt{LxCat} database (gas) and Ref.~\cite{bib:boyle} (liquid) and single-differential cross sections from Ref.~\cite{bib:WT} (solid lines).}
    \label{fig:diffxsec}
\end{figure}

In its default configuration, the simulation follows the treatment of Ref.~\cite{bib:WT} which implements energy-dependent cross sections and isotropic momentum-scattering interactions. For inelastic interactions cross sections are available as a function of the electron's scattering angle. For interactions in gas, double-differential cross sections are available from the \texttt{LxCat} database. \textcolor{black}{In liquid, where the electron's de Broglie wavelength is large compared to the typical separation between argon atom, a model of individual collisions with nearby atoms starts to break down (as introduced when discussing the Cohen-Lekner theory). This makes the inclusion of differential cross sections accounting for angular deflections more complex. Recent work by Boyle et al. in Ref.~\cite{bib:boyle} provides  a generalization of the Cohen-Lekner theory which includes details of the anisotropy of the effective momentum and energy-transfer interactions introduced by Cohen and Lekner. The differential cross sections for elastic scattering in liquid as a function of both energy and angle can be seen in Fig. 10 of Ref.~\cite{bib:boyle} by Boyle et al. Theoretical calculations for double-differential energy-transfer and momentum-transfer cross sections implemented by Boyle et al. remove several assumptions used by Cohen and Lekner, something now possible thanks to advances in computational methods. This development allowed for the more detailed differential cross section calculations reported in their work.} The simulation has been expanded to implement the energy and angle-dependent scattering available from these double-differential cross section inputs. The integrated 1D energy-dependent cross sections are compared in figure~\ref{fig:diffxsec}, showing minor overall changes. Results presented in this work focus on leveraging the single-differential energy-dependent cross section model, currently set as the default. Future work will expand the use of differential cross section on argon in the study of specific microphysics processes where a more accurate modeling of the angular distribution of scattered electrons can provide significant impact in modeling accuracy.

\section{Microphysics Simulation}
\label{sec:simulation}

This chapter describes how individual electrons are tracked and allowed to interact and propagate in 3D under the influence of a (variable) electric field in the simulation. The cross sections described in Chapter~\ref{sec:xsec} are the core input to the full simulation, which is carried out through a 
Monte Carlo time-domain simulation of individual electron-argon interactions. The null-collision technique is utilized to simulate interactions. In this approach, a constant interaction rate $K$ is computed such that $K > \sigma_{\rm tot}\left(E\right) v$ with $v$ the magnitude of the electron's velocity and $\sigma_{\rm tot}\left(E\right)$ the total cross section at any given energy. A time-step $t$ is then computed at each iteration in the simulation, sampling from an exponential distribution with constant
\begin{equation}
\label{eq:tau}
   \tau = \frac{K}{ 2 n},
\end{equation}
where \textcolor{black}{$n$ is the argon number density and} the factor of two aims to increase the sampling rate. The maximum time-step simulated is truncated at $3\tau$.

Given a simulated time-step $t$, the electron\textcolor{black}{'s position, velocity,
and acceleration} are updated classically under the action of the local electric field (denoted as $\mathbf{\nabla}V$):
\begin{eqnarray}
\vec{x}_k &=& \vec{v}_{k-1} t + \frac{1}{2} \vec{a}_{k-1}  t^2, \\
\vec{v}_k &\mathrel{+}=& \vec{a}_{k-1} t, \\
\vec{a}_k &=& \frac{e}{m_e} \vec{\mathbf{\nabla}} V\left(\vec{x}_{k}\right),
\end{eqnarray}

\begin{figure}%[ht]
\begin{center}
  \includegraphics[width=0.99\textwidth]{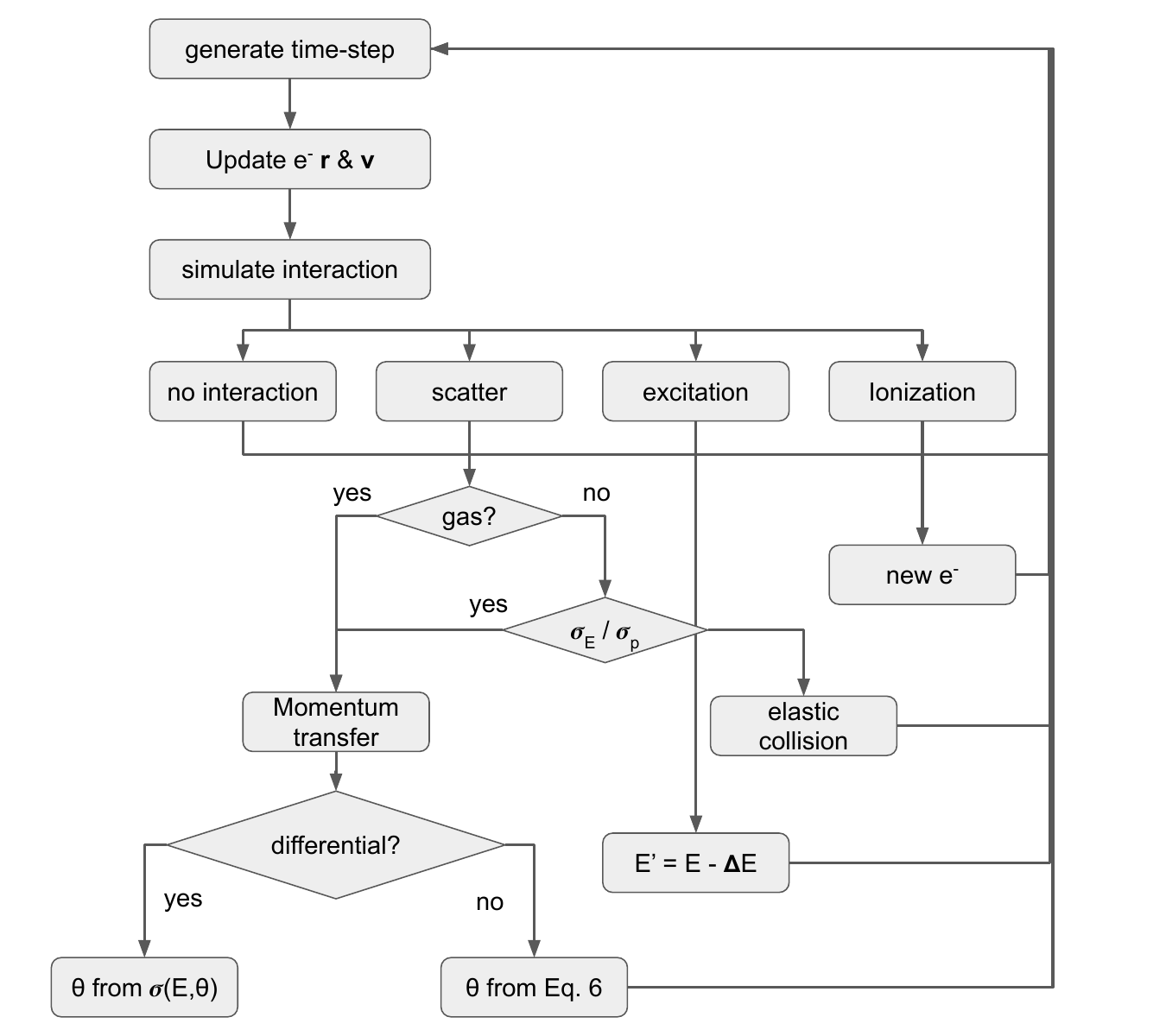}
  %\captionsetup{margin=0.1cm}
  \caption{Simulation flow-chart showing the logic followed at each step of an electron's trajectory simulation. Once a time-step is simulated, the electron's position is updated and a possible interaction mode is simulated. Depending on the interaction type, whether the particle is propagating in liquid or gas, and whether differential cross sections are used or not, the electron's energy, velocity, and direction are updated as described by the flow-chart diagram.}
  \label{fig:chart}
  \end{center}
\end{figure}

\textcolor{black}{where $e$ is the electron's charge.} After propagating the electron in space, a decision on its possible interaction with an argon atom is made. For each interaction mode defined in Sec.~\ref{sec:xsec} and labeled as $i$, the interaction probability is computed as:
\begin{equation}
    p_i = \frac{\nu \sigma_i\left(E\right)}{K},
\end{equation}
\textcolor{black}{where $\nu$ is the electron's velocity.} The sum of all values of $p_i$ for the various interaction modes is guaranteed to be less than one due to how the quantity $K$ is constructed. A random number in the range $0-1$ is drawn and an interaction is simulated based on each interaction mode's probability, allowing for null interactions. At this stage, the energy and velocity vector of the electron are updated depending on the interaction mode. 
For inelastic interactions, the speed and direction of scattering electrons are updated following the treatment of Wojcik and Tachiya:

\begin{equation}
    \vec{\textbf{v}}' = \frac{M u}{m + M}\vec{\textbf{n}} + \frac{m\vec{\textbf{v}} + M\vec{\textbf{v}_M}}{m + M}
\end{equation}
With $\vec{\textbf{n}}$ a randomly drawn unit vector and $u$ a randomly drawn value of the argon atom's velocity based on the Boltzmann velocity distribution at liquid argon temperature. If the collision is treated as an energy-transfer process, only the magnitude of the velocity is updated, but not the direction.
A flow-chart summarizing the simulation is presented in Figure~\ref{fig:chart}.

%\subsection{Electron Transport}

\begin{figure}%[H]
\begin{center}
  \includegraphics[width=0.6\textwidth]{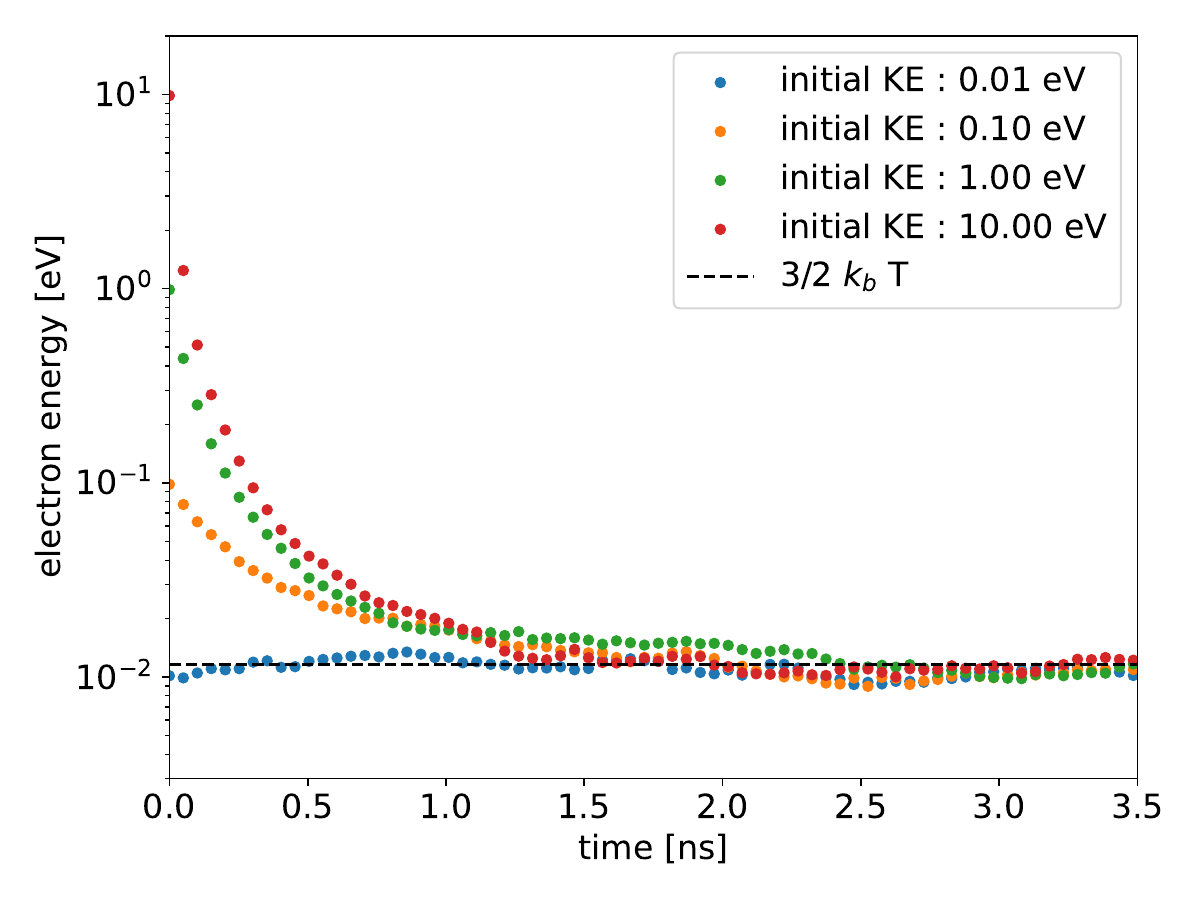}
  %\captionsetup{margin=0.1cm}
  \caption{Thermalization of electrons at different initial energies simulated in the absence of an electric field. All electrons recover their characteristic thermal energy $3/2 k_b T$ after a few ns, consistent with the results of Wojcik and Tachiya~\cite{bib:WT}.}
  \label{fig:thermalization}
  \end{center}
\end{figure}

A first evaluation of the simulation is performed by propagating electrons of different initial energy without the presence of an electric field. Figure~\ref{fig:thermalization} shows the evolution in time of the median energy of simulated populations of 100 electrons produced with initial energies f 0.01, 0.1, 1, and 10 eV. All populations quickly thermalize through interactions with argon atoms, reaching a stable energy comparable to $3/2 k_B T$ with T the temperature of liquid argon and $k_B$ the Boltmann constant. The time-scale for thermalization is of the order of $\mathcal{O}$(1 ns), consistent with the results of Wojcik and Tachiya~\cite{bib:WT}.

A second demonstration of the propagation of electrons in the simulation is shown in figure~\ref{fig:garcascade}, which presents the trajectory of individual electrons simulated in a liquid argon medium under the influence of different electric fields. The four panels, simulated at 0, 1, 3, and 10 kV/cm, show the change in behavior of the electron clouds at different values of the electric field. The increase in distance propagated in the direction of the electric field gives a first qualitative indication of the field-dependence of the electron drift velocity, which will be studied in more detail in Ch.~\ref{sec:swarm}
\newpage
\begin{figure}%[H]
    \centering
    \begin{subfigure}[b]{0.85\textwidth}
    \centering
    \includegraphics[width=1.00\textwidth]{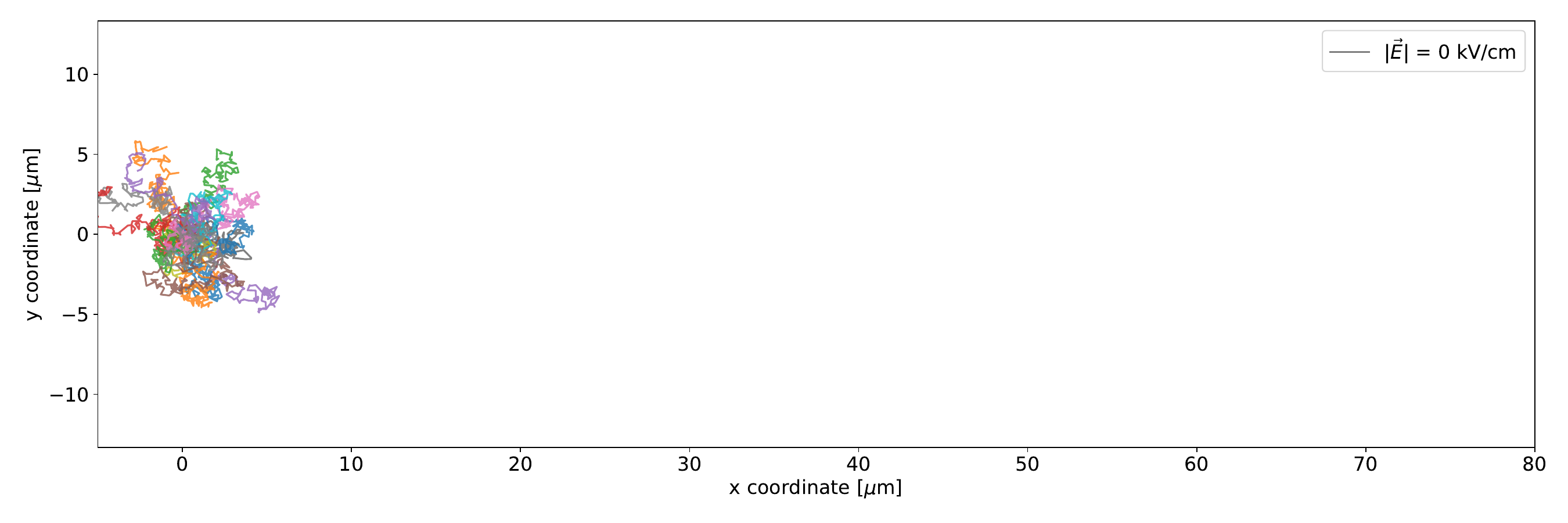}
    \caption{0 kV/cm}
    \end{subfigure}
    \begin{subfigure}[b]{0.85\textwidth}
    \centering
    \includegraphics[width=1.00\textwidth]{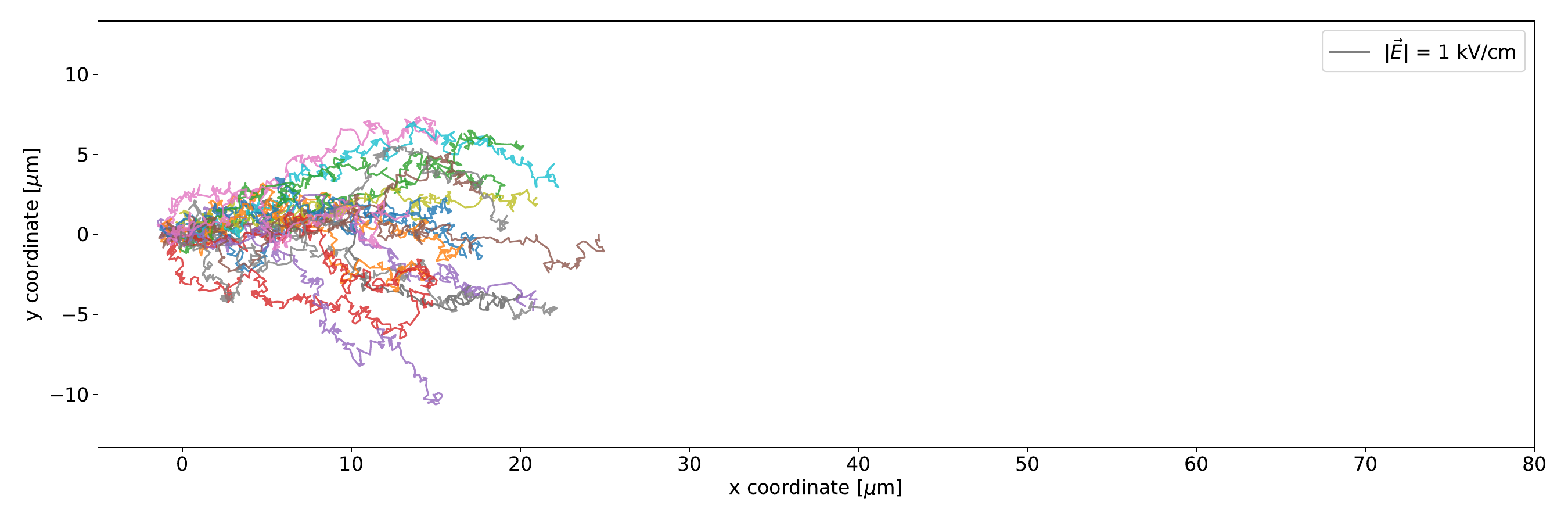}
    \caption{1 kV/cm}
    \end{subfigure}
    \begin{subfigure}[b]{0.85\textwidth}
    \centering
    \includegraphics[width=1.00\textwidth]{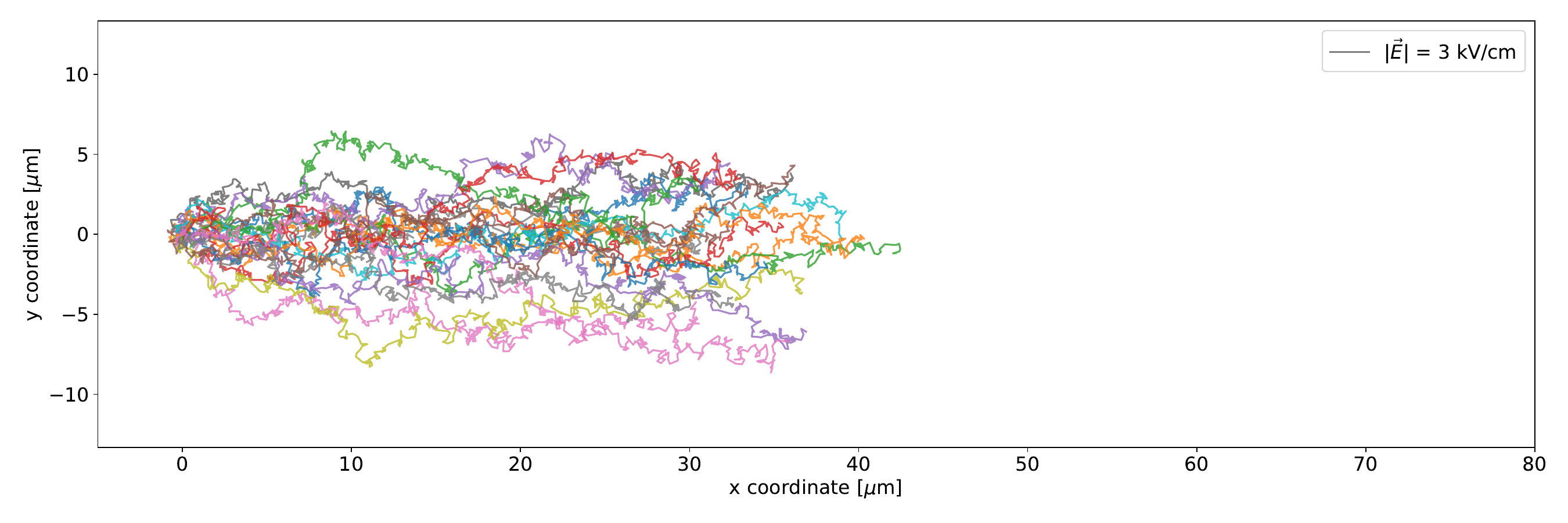}
    \caption{3 kV/cm}
    \end{subfigure}
    \begin{subfigure}[b]{0.85\textwidth}
    \centering
    \includegraphics[width=1.00\textwidth]{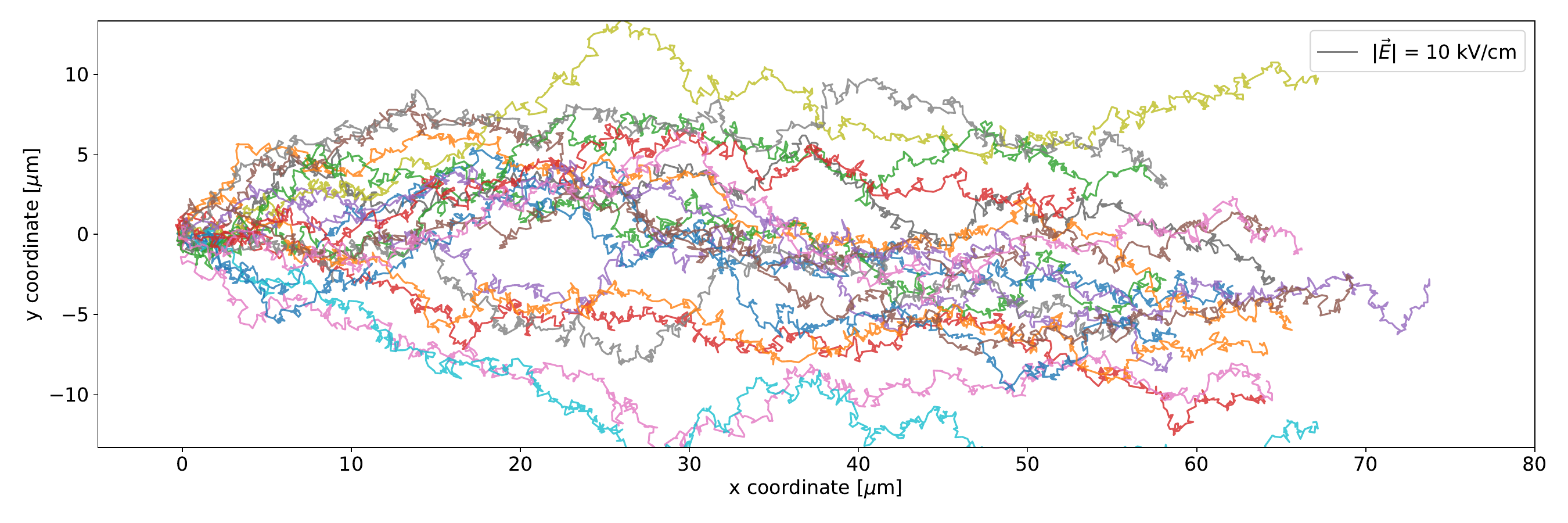}
    \caption{10 kV/cm}
    \end{subfigure}
    \caption{Trajectories of several electrons propagating in uniform electric fields of different strength (0, 1, 3, and 10 kV/cm) oriented in the positive $x$-coordinate.}
    \label{fig:trajectories}
\end{figure}
\newpage

Finally, figure~\ref{fig:garcascade} presents the simulation of an electron avalanche in gaseous argon at 2 psi at a field of 5 kV/cm. This simulation is obtained by tracking the primary electron and each secondary electron produced via ionization. Each ionization electron is propagated independently, with an initial position coinciding with the location of the ionization interaction.

\begin{figure}%[H]
\begin{center}
  \includegraphics[width=0.95\textwidth]{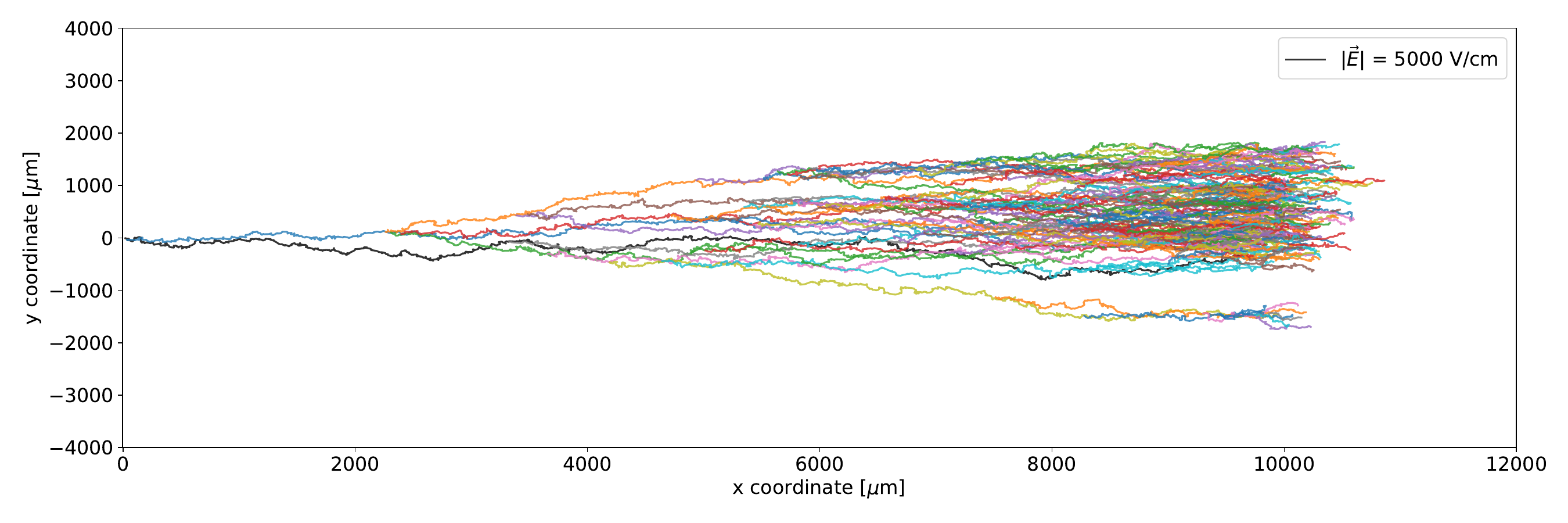}
  %\includegraphics[width=0.25\textwidth]{avalanchelar.png}
  %\captionsetup{margin=0.1cm}
  \caption{Electron cascade in GAr. Each ionization electron is tracked with a different color. Branches in the cascade indicate points where an ionization occurs, producing a new free electron which is independently tracked in the simulation.}
  \label{fig:garcascade}
  \end{center}
\end{figure}

\section{Simulation Validation}
\label{sec:swarm}
This chapter presents validations of the simulation via comparisons to experimental measurements of the macroscopic swarm parameters of drift velocity and diffusion in gaseous and liquid argon. This aims to validate the simulation's performance at electron energies below $\mathcal{O}$(10) eV, where elastic scattering dominates. Validations of the simulation at higher energies for the modeling of ionization are presented in Chapter~\ref{sec:ionization} where studies of ionization are carried out. 

\begin{figure}%[H]
\begin{center}
  \includegraphics[width=0.95\textwidth]{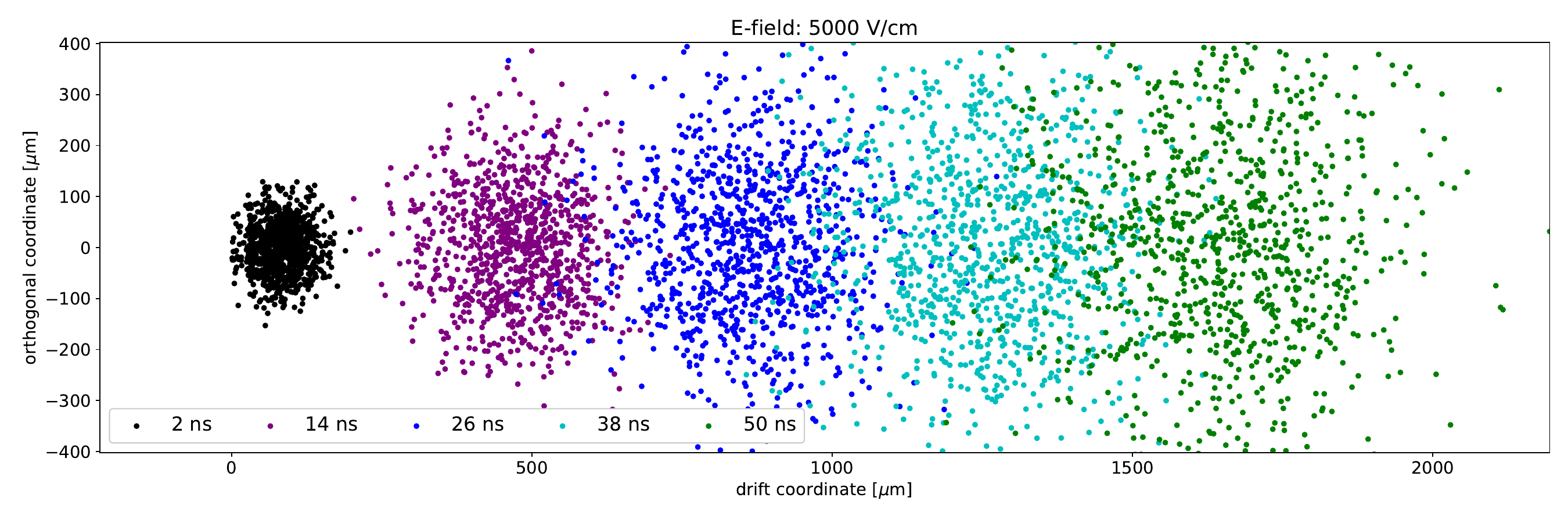}
  %\captionsetup{margin=0.1cm}
  \caption{Example electron cloud of $10^3$ electrons propagating in gaseous argon at 2 PSI under the influence of a 5 kV/cm electric field. The image shows snapshots of the electron cloud after 2, 14, 26, 38, and 50 ns.}
  \label{fig:swarm}
  \end{center}
\end{figure}

Datasets for drift velocity and diffusion parameters in gas and liquid used to benchmark the simulation are obtained from tabulated \texttt{LxCAT} databases~\cite{bib:lxcat}. For the simulation itself, these quantities are computed by propagating swarms of $\mathcal{O}(10^2) \textrm{-} \mathcal{O}(10^3)$ electrons for $\mathcal{O}(10^{-9}) \textrm{-} \mathcal{O}(10^{-6})$ seconds under different $E$-field configurations and computing the collective behavior of the electron cloud. 
%The macroscopic swarm parameters that are tested are the electron drift velocity and diffusion coefficients.
An example showing the collective behavior of $10^3$ electrons propagating through the simulation is shown in figure~\ref{fig:swarm}.

\subsection{Drift Velocity}
The drift velocity is calculated by computing the average distance traveled by the simulated electrons in the drift direction $x$ over the total time for which the simulation is allowed to run. 

\begin{figure}%[H]
\begin{center}
  \includegraphics[width=0.5\textwidth]{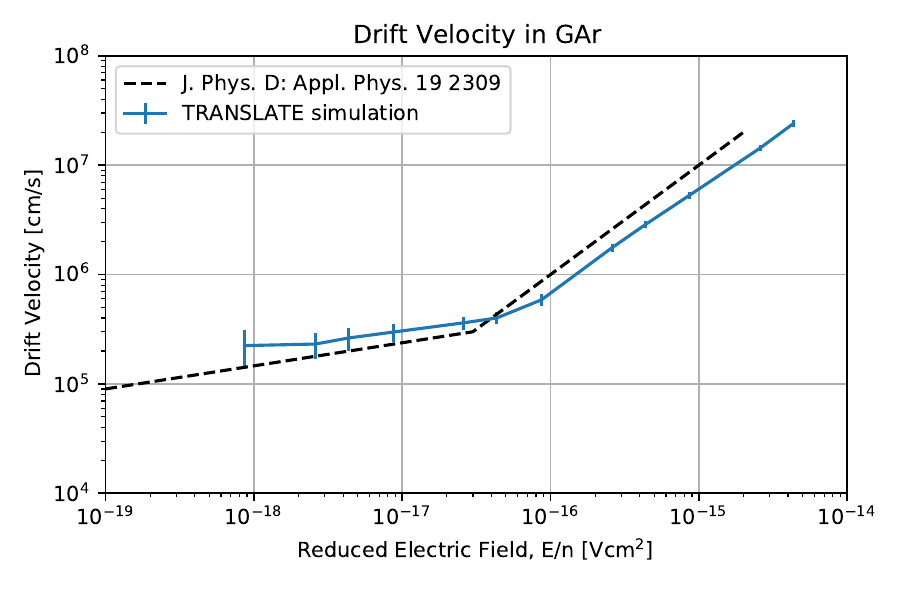}
  \includegraphics[width=0.45\textwidth]{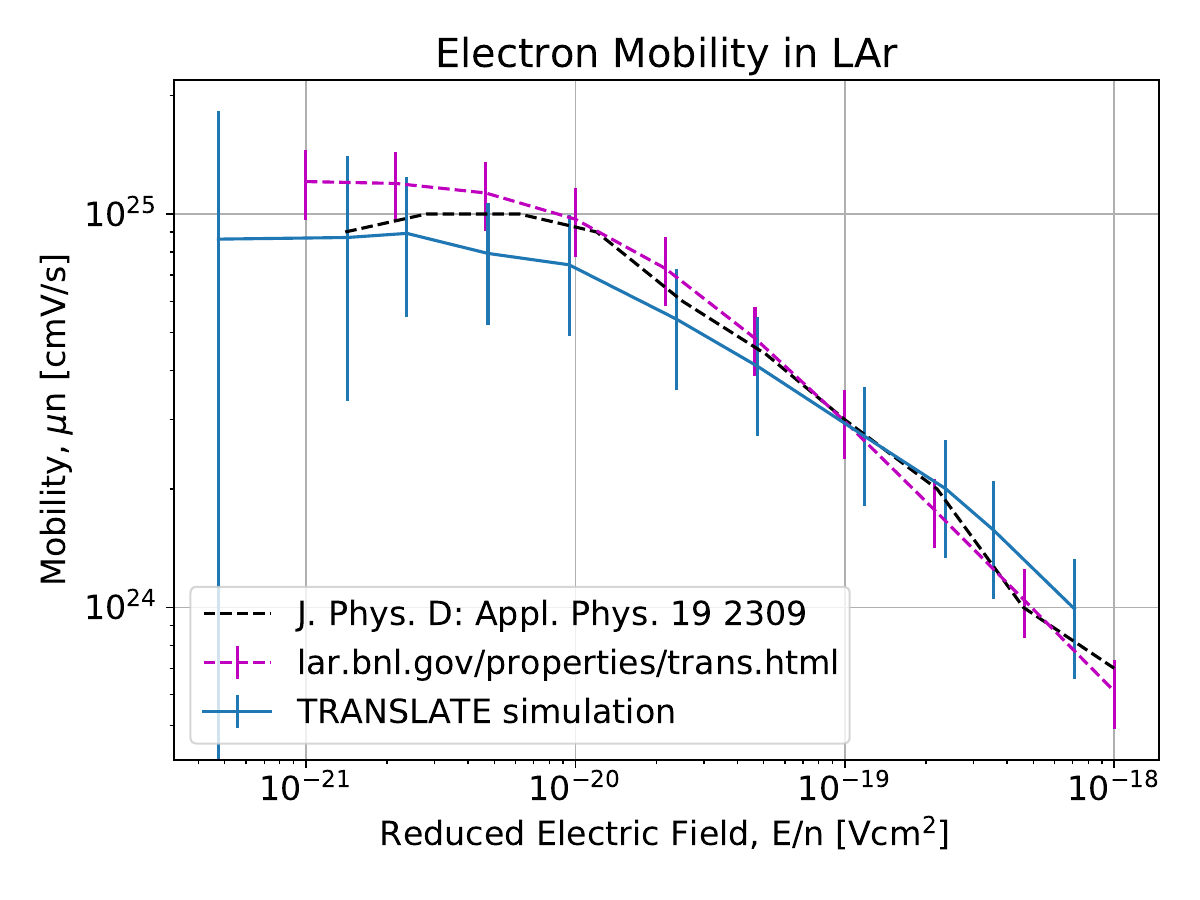}
  %\captionsetup{margin=0.1cm}
  \caption{Validation of drift velocity and electron mobility in gaseous (left) and liquid argon. Comparisons for the drift velocity in gaseous argon come from Ref.~\cite{bib:puech}. For LAr, the experimental values compared to come from Refs.~\cite{bib:puech,bib:bnllarproperties}.}
  \label{fig:driftvel}
  \end{center}
\end{figure}

The electron drift velocity and diffusion in argon are benchmarked against datasets obtained from the collection of electron swarm parameters available on the \texttt{LxCat} database~\cite{bib:lxcat}. The drift velocity computed in the simulation is shown in figure~\ref{fig:driftvel}. The left panel shows the drift velocity in gas, while the right image shows the electron mobility in liquid (electron mobility is given by the drift velocity divided by the magnitude of the electric field). Both are shown as a function of the reduced electric field, computed as the electric field strength divided by the argon atom number density in units of Townsends ($1 \textrm{Td} = 10^{-21} \textrm{V}\times \textrm{m}^2$). The drift velocity is converted to units of $\textrm{m}^{-1} \textrm{s}^{-1}$ by multiplying by the gas density, to reflect the units of swarm parameters as described in the \texttt{LxCAT} database. Both gas and liquid show good agreement between the simulation and data, with the simulation tracking the kink in drift velocity expected in gas. \textcolor{black}{This kink is a result of the change in the cross section at $\sim10$ eV, where the elastic cross section begins to decrease and inelastic processes become relevant}. Over the span of three orders of magnitude in electric field strength, the simulation is tracking the collective behavior of electrons in gaseous argon in these two metrics within $\mathcal{O}$(20\%) for most calculated data points.

\subsection{Electron Diffusion}
\begin{figure}%[H]
\begin{center}
  \includegraphics[width=0.98\textwidth]{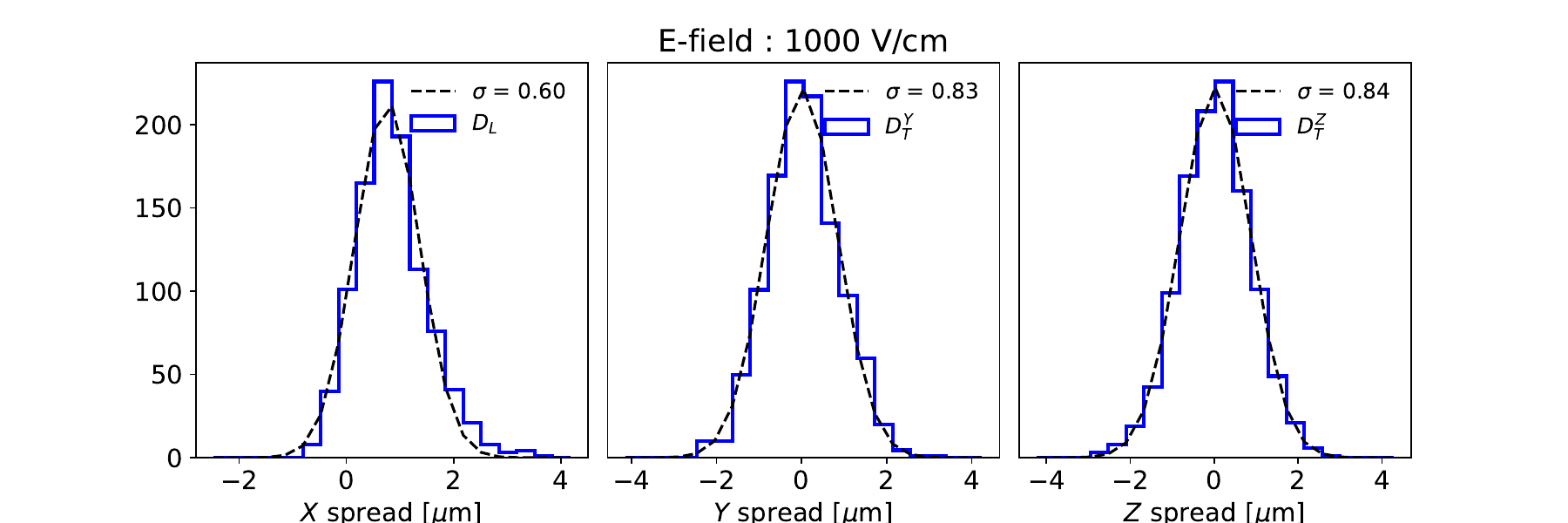}
  %\captionsetup{margin=0.1cm}
  \caption{Spread in electron cloud for an example simulation, and Gaussian fit used to extract diffusion coefficient in all three coordinates.}
  \label{fig:diffusioncloud}
  \end{center}
\end{figure}

Electron diffusion is measured  through an analysis of the electron clouds after they have been propagated through the simulation for a specified amount of time. The analysis is carried out similarly to the analysis presented in Ref.~\cite{bib:diffusionMicroBooNE} and is briefly summarized below. The spread in the simulated electron cloud along the longitudinal ($x$) and transverse ($y$, and $z$) coordinates is fit to a Gaussian function, and the width of the distribution is used to extract the diffusion coefficient. Examples of electron transverse and longitudinal cloud widths, and the Gaussian fit to their distribution, are shown in figure~\ref{fig:diffusioncloud}. The longitudinal and transverse diffusion coefficients ($D_L$ and $D_T$ respectively) are calculated as $D = \sigma^2 / (2 t_{\rm drift})$, with the electron cloud spread in the drift coordinate, $\sigma_x$, used to calculate the longitudinal diffusion coefficient $D_L$, and the spread in $y$ and $z$ to compute $D_T$.

\begin{comment}
\begin{figure}%[H]
    \centering
    \begin{subfigure}[b]{0.25\textwidth}
    \centering
    \includegraphics[width=1.00\textwidth]{dl_field_30.pdf}
    \caption{$\sigma_L$ at 30 V/cm}
    \end{subfigure}
    \begin{subfigure}[b]{0.25\textwidth}
    \centering
    \includegraphics[width=1.00\textwidth]{dt_y_field_3000.pdf}
    \caption{$\sigma_T$ at 3 kV/cm}
    \end{subfigure}
    \begin{subfigure}[b]{0.25\textwidth}
    \centering
    \includegraphics[width=1.00\textwidth]{dt_field_10000.pdf}
    \caption{$\sigma_T$ at 10 kV/cm}
    \end{subfigure}
    \caption{Trajectories.}
    \label{fig:diffusionfits}
\end{figure}
\end{comment}

\begin{figure}%[H]
\begin{center}
  \includegraphics[width=0.45\textwidth]{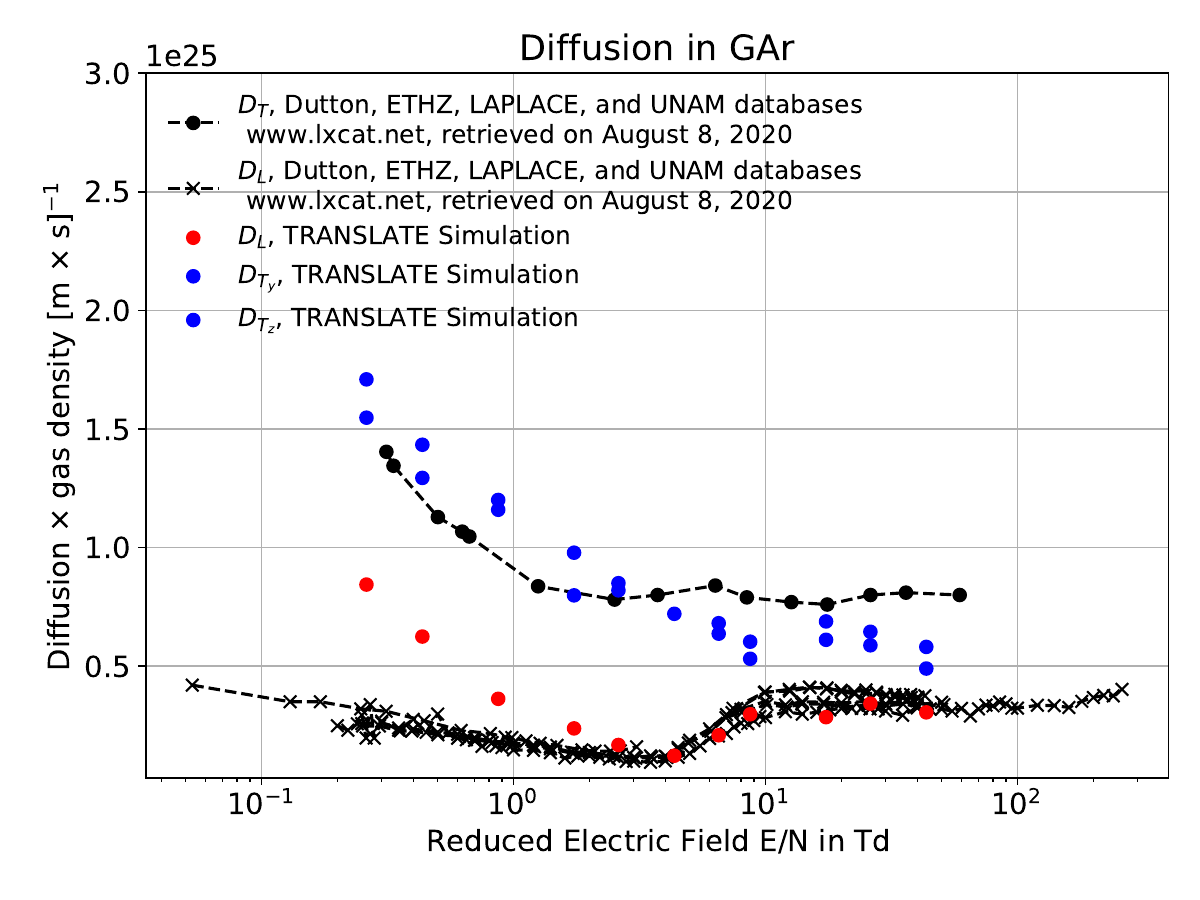}
  \includegraphics[width=0.45\textwidth]{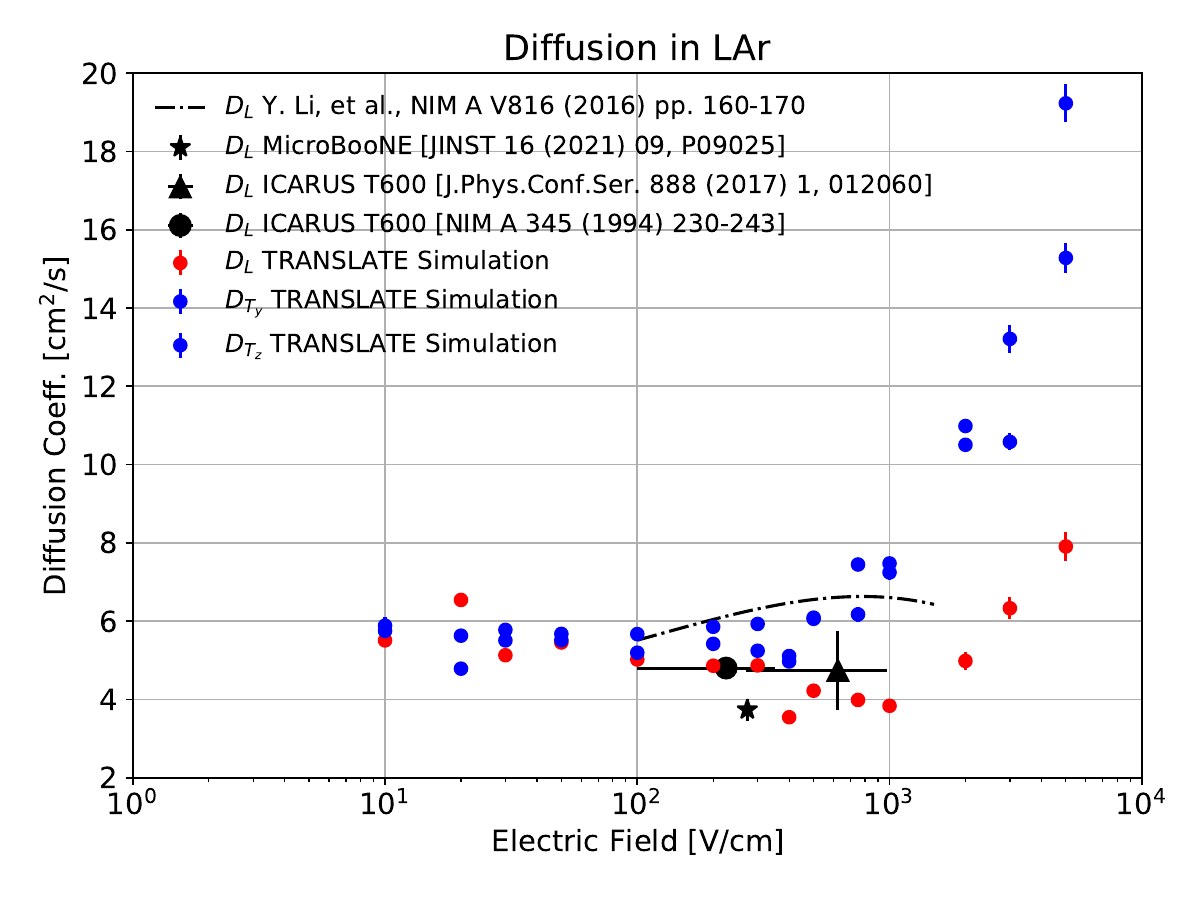}
  %\captionsetup{margin=0.1cm}
  \caption{Transverse ($D_T$) and longitudinal ($D_L$) diffusion coefficients calculated in gas and liquid argon from the \texttt{TRANSLATE} simulation output, compared to available world data.}
  \label{fig:gardiffusion}
  \end{center}
\end{figure}

Results of the measurement of electron diffusion from the simulation are presented in Fig.~\ref{fig:gardiffusion} for both gas and liquid, and compared to available datasets. 
Comparisons in gas are made to datasets available from the \texttt{LxCat} database~\cite{bib:lxcat}, specifically Ref.~\cite{bib:DUTTON,bib:ETHZ,bib:LAPLACE,bib:UNAM}. Diffusion coefficients are found to be in good agreement with data, where available. In gas, the coefficient of longitudinal diffusion slightly over (under) estimates the data at low (high) electric fields, but follows the expected trend. Longitudinal diffusion data is from the UNAM database~\cite{bib:UNAM} which cites a 10\% uncertainty on the measurement, of order the differences seen with respect to the simulation.
The modeling of transverse diffusion in gas also matches well the measurement, especially at reduced electric fields above a few Townsend, including an accurate reproduction of the kink seen between 4 and 10 Td. At low values of the reduced electric field, corresponding to $\mathcal{O}$(100 V) for a pressure of 2 PSI, the simulation begins to deviate from measurements, over-estimating $D_L$. We hypothesise that this can be attributed to the simulation environment omitting any impurities, present in some residual amount in experimental setups used for the referenced measurements, which causes a reduction of the diffusion coefficient.

Comparisons in liquid are performed with some of the data measurements available from the ICARUS~\cite{bib:diffusionICARUS1,bib:diffusionICARUS2} and MicroBooNE~\cite{bib:diffusionMicroBooNE} experiments, as well as the results of Ref.~\cite{bib:BNLdiffusion}, and show good agreement with the simulation with values of $D_L$ of $4\textrm{--}5$ cm${}^2$/sec.
Finally, the upturn in $D_L$ observed above $10^3$ V/cm is consistent with the expectation from the Atrazhev-Timoshkin theory~\cite{bib:diffusionAT} as described in Ref.~\cite{bib:diffusionMicroBooNE}.

\section{Studies of Electron Amplification in Argon}
\label{sec:ionization}

Proportional electron amplification in strong electric fields is a common technique employed to enable sufficient gain in order to amplify and collect ionization signatures from small energy deposits. In noble element detectors this amplification is often carried out via the use of GEM detectors in gas with local electric fields of $\mathcal{O}$(kV/cm) field-strengths. 

Electron amplification in liquid can provide a new path towards developing low-threshold detectors sensitive to primary ionization electrons produced in low-energy interactions without the need to interface a gaseous layer in dual-phase TPCs. Achieving amplification in liquid noble elements such as argon and xenon has been proposed and studied in several efforts~\cite{bib:larmult01,bib:larmult02,bib:larmult03,bib:larmult04} but has not yet yielded stable performance that can be utilized for low-energy physics measurements. LArCADe (Liquid Argon Charge Amplification Devices) is an ongoing experimental effort that pursues this instrumentation development. Part of the motivation for the simulation package developed in this work is to be able to accurately model the behavior of electron propagation at high electric fields in order to study the necessary conditions for obtaining amplification in liquid argon and thus guide experimental work on the matter.

The primary challenge for achieving electron multiplication in liquid is given by the need to accelerate drifting electrons to a high enough energy such that they can induce secondary ionization before scattering off of an argon atom, slowing down in the process. In liquid, owning to the much larger density, this requires significantly stronger fields compared to gas.
The nature of scattering cross sections which grow as the electron's energy approaches $\mathcal{O}$(10) eV makes this an ``uphill battle'', as the electron is more likely to scatter and lose energy as it approaches the threshold for ionization. The field strength required for ionization in liquid is $\mathcal{O}(10^6)$ V/cm, roughly three orders of magnitude larger than in gas, proportional to the ratio in densities.% roughly proportional to the 
%To get a sense of scale for the process, the requirement on the local electric field $\vec{E}$ needed to accelerate electrons in order to meet this condition must be such that an electron gains $\mathcal{O}$(10) eV of kinetic energy within one interaction length:
%\begin{eqnarray}
%\vert \vec{E} \vert \lambda &=& \unit[10]{eV}   \\
%\lambda &=& \frac{1}{n \sigma}
%\end{eqnarray}
%In liquid, $\sigma \sim \unit[10^{-15}]{cm^{2}}$, $n = \unit[10^{22}]{cm^{-3}}$. Solving for $\vec{E}$, this demands an electric field strength of $\sim\unit[10^8]{V/cm}$.

\begin{figure}%[H]
\begin{center}
  \includegraphics[width=0.49\textwidth]{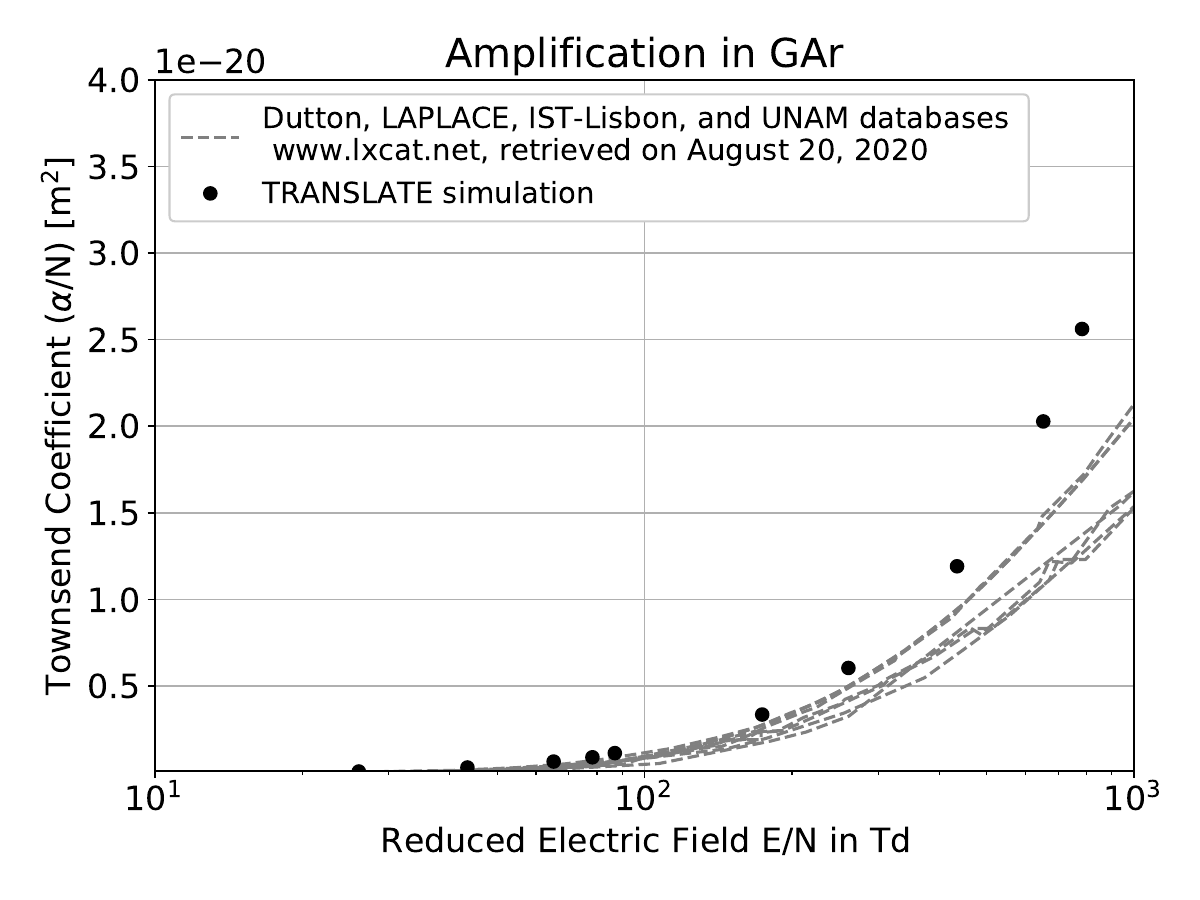}
  \includegraphics[width=0.49\textwidth]{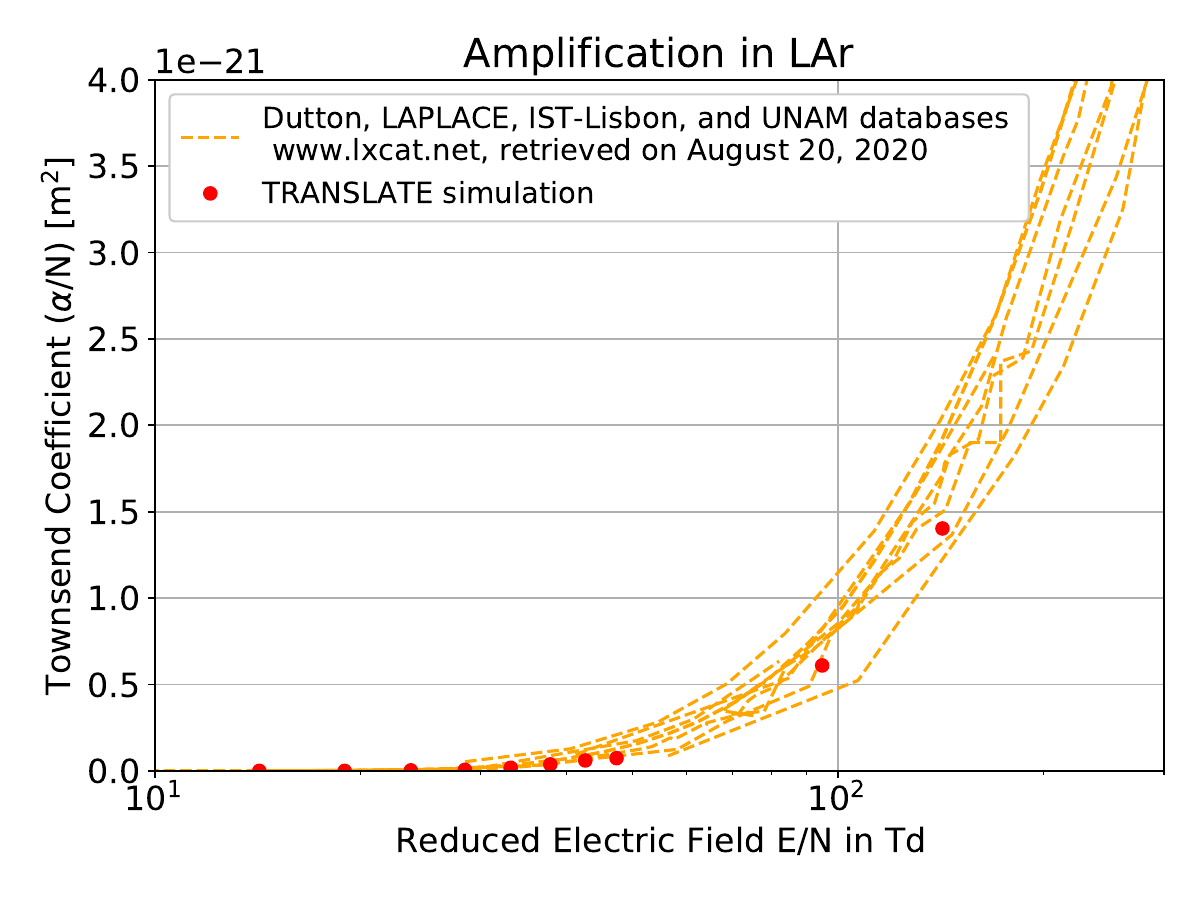}
  %\captionsetup{margin=0.1cm}
  \caption{Townsend coefficient, measuring charge amplification, in GAr (left) and LAr (right). Points are from the \texttt{TRANSLATE} simulation, while the dotted curves are from experimental results fro gaseous argon.}
  \label{fig:townsend}
  \end{center}
\end{figure}

\texttt{TRANSLATE} allows for the tracking and propagation of secondary electrons induced by ionization. This mode of running the simulation, however, adds significant run time due to the large number of electrons to track. To calculate the Townsend coefficient and gain amplification factor in the simulation, we instead propagate a single electron, count the individual ionizations this electron induces, and assume exponential binomial growth to estimate the total amplification electrons as $2^{n}$ with $n$ the ionization electrons induced by the initial electron tracked in the simulation.
The measurement of the Townsend coefficient $\alpha$ in gas and argon is shown in figure~\ref{fig:townsend} compared to data obtained from the \texttt{LxCat} database. The Townsend coefficient $\alpha$ is defined as
\begin{equation}
   \alpha = \ln\left(\frac{N}{N_0}\right) \frac{1}{d}
\end{equation}
with $N$ and $N_0$ the number of ionization electrons after and at the start of their transport, respectively, and $d$ the distance traveled.

The results show good agreement between experimental data and the simulation, with deviations between the two of order the variance between different datasets from the \texttt{LxCat} database. When expressed in units of Townsend (which normalize by the density) the gas and liquid results are consistent owing to the similarity of the cross section models. Expressed in units of absolute electric field this translates to a difference in the onset of multiplication of several orders of magnitude, comparable to the density difference between the two fluids. 
\begin{figure}%[H]
\begin{center}
  \includegraphics[width=0.5\textwidth]{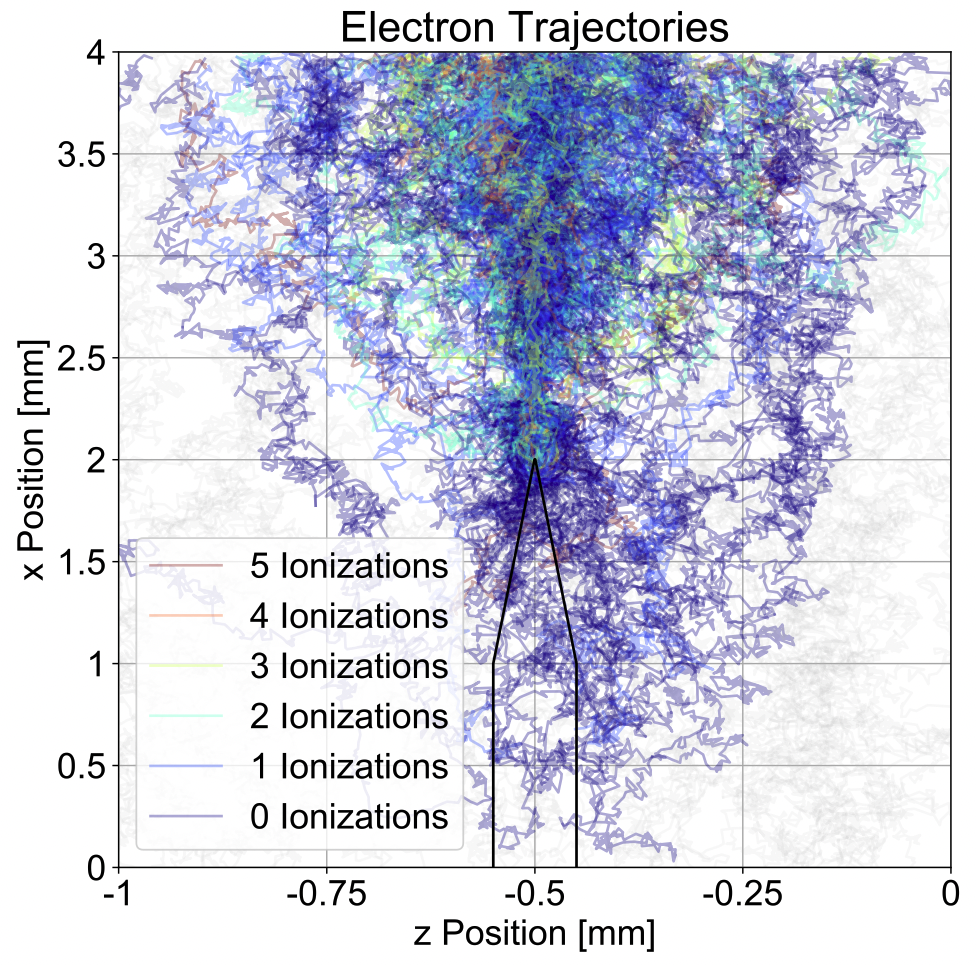}
  \caption{Simulated electron trajectories in the presence of a locally varying electic field in the vicinity of a conducting ``tip''.}
  \label{fig:tip}
  \end{center}
\end{figure}
One method for achieving electric fields of the magnitude necessary to induce ionization in liquid is to modify the local geometry near the charge sensors in order to locally induce conditions that lead to charge amplification. Tip geometries, such as the ones proposed and tested in Ref.~\cite{bib:larmult01,bib:larmult02,bib:larmult03,bib:larmult04}, can provide such conditions. Modeling the behavior of electron swarms in this environment requires the ability to simulate position-dependent electric field conditions. This is something that has been incorporated in the \texttt{TRANSLATE} simulation. Figure~\ref{fig:tip} shows an example of electron trajectories simulated in a locally varying electric field in the presence of a micron-scale tip. Electric field lines converge on the tip, guiding the path of drifting electrons, and causing some to produce secondary ionization electrons. This image is presented as an illustration of the flexibility of the simulation package developed. %\sout{The electric field map in the presence of the tip is generated with the \texttt{COMSOL} simulation package, which is then imported in the \texttt{TRANSLATE} software in order to simulate electron propagation.}
\textcolor{black}{The implementation of position-dependent E-fields was achieved by passing as input simulation maps constructed with the \texttt{COMSOL}\textregistered~\cite{bib:COMSOL} simulation package. The electic field at any given position $\vec{E}(x,y,z)$ is the computed in \texttt{TRANSLATE} by interpolating between values provided by the input E-field map. While microscopic effects due to detector geometry are accounted for in the \texttt{COMSOL}\textregistered simulation, the impact of molecular dynamics induced by the argon medium are not.}

%The predicted number of ionizations per unit distance in liquid according to the simulation is shown in figure~\ref{fig:laramplification}. A turn-on in the amount of amplification seen is noticeable at fields of $\mathcal{O}\left(5\times10^6\right)$ V/cm.

\begin{comment}
\begin{figure}%[H]
\begin{center}
  \includegraphics[width=0.6\textwidth]{lar_amplification.pdf}
  %\captionsetup{margin=0.1cm}
  \caption{Secondary ionization in liquid argon \textcolor{red}{add more data-points}.}
  \label{fig:laramplification}
  \end{center}
\end{figure}
\end{comment}

\section{Conclusions}
\label{sec:conclusions}
\par We have developed a simulation package aimed at modeling the transport of electrons in argon over a broad range of electric field strengths. This simulation, leveraging cross section calculations and measurements available from the literature, expands on the simulation developed by Wojcik and Tachiya in Ref.~\cite{bib:WT} to include the processes of argon excitation and ionization. The simulation tracks individual electrons in 3D, allowing for the possibility of implementing position-dependent detector geometries and electric field-maps. We benchmark the performance of the simulation by comparing measured swarm parameters of drift velocity and electron diffusion with data from gaseous and liquid argon, and electron amplification for gaseous argon. The simulation is then used to quantify the conditions necessary to trigger electron amplification in liquid. Together with this article, the \texttt{TRANSLATE} code package is made available for future development and expanded uses. We in particular highlight the possibility of additional microphysics studies of electron-ion recombination, quenching, diffusion, and light production which can be produced with an expanded simulation that incorporates cross sections with $\textrm{Ar}^{+}$ ions and molecular impurities.

\section{Acknowledgements}

This work was supported in part by Fermilab's LDRD support for the LArCADe project. Zack Beever was supported through the Fermilab Community College Internships (CCI) program. Preparation of this manuscript was in part supported by the U.S. Department of Energy, Office of Science. We are grateful to Daniel Cocks and Gregory Boyle for useful communication on electron-argon cross sections and for providing tabulated differential cross sections for electron scattering in gaseous argon.





\end{document}